\documentclass[%
 reprint,
 amsmath,amssymb,
 aps,
]{revtex4-2}
\usepackage{multirow}
\usepackage{color}
\usepackage{graphicx}
\usepackage{dcolumn}
\usepackage{bm}
\usepackage{hyperref}
\usepackage{xcolor}

\begin{document}

\preprint{APS/123-QED}

\title{Thermodynamics of the  Isospectral family of holographic vector mesons}

\author{Miguel Angel Martin Contreras}
\email{miguelangel.martin@usc.edu.cn}
\affiliation{
 School of Nuclear Science  and Technology\\
 University of South China\\
 Hengyang, China\\
 No 28, West Changsheng Road, Hengyang City, Hunan Province, China.
}


\author{Alfredo Vega}%
 \email{alfredo.vega@uv.cl}
\affiliation{%
 Instituto de F\'isica y Astronom\'ia, \\
 Universidad de Valpara\'iso,\\
 A. Gran Breta\~na 1111, Valpara\'iso, Chile
}

\author{Saulo Diles}
\email{smdiles@ufpa.br}
\affiliation{Campus Salin\'opolis,\\ Universidade Federal do Par\'a,\\
68721-000, Salin\'opolis, Par\'a, Brazil}
\affiliation{Unidade Acadêmica de Física,\\ Univ. Federal de Campina Grande, R. Aprígio Veloso, 58429-900 - Campina Grande}

\begin{abstract}
We study the thermal behavior of the $\rho$ meson using the isospectral family of the softwall AdS/QCD model. By computing spectral functions at finite temperature and chemical potential for different members of this family, we isolate the effect of the ground-state electromagnetic decay constant $f_1$ on the melting temperature $T_m$ of the $\rho(770)$ meson. A clear monotonic increase of $T_m$ with $f_1$ is found, supporting the interpretation of $f_1$ as a key scale controlling quarkonium dissociation. For excited states, the same qualitative trend appears but is strongly suppressed as the radial quantum number increases. Using the isospectral parameter to fix $f_1$ to its experimental value ($226$ MeV) yields a holographic model whose spectral function gives a melting temperature $T_m = 157$ MeV and a smooth crossover from confinement to deconfinement. The thermal mass shows a mild decrease near the critical point, while the width grows monotonically. Our results demonstrate that the isospectral transformation provides a controlled way to adjust ground-state decay constants without altering the mass spectrum, enabling precise studies of medium effects on vector mesons.
\end{abstract}

\maketitle

\section{Introduction}
Supersymmetric quantum mechanics revived an old topic in the theory of ordinary differential equations known as the Darboux transformation \cite{darboux1882proposition, Witten:1981nf}. For a given Schrödinger operator with a discrete spectrum, there exists a one-parameter family of operators sharing the same eigenvalues but having different eigenfunctions. Such a transformation provides a systematic method for constructing equivalent potentials that define a family of \emph{isospectral Schrödinger operators}. 

The isospectral transformation has been explored in the context of bottom-up holographic QCD models. These models arise from the AdS/CFT correspondence \cite{Maldacena:1997re} and describe the hadronic spectrum via holographic mapping. In bottom-up AdS/QCD, the bulk modes representing confined quark states (hadrons) are eigenfunctions of a Schrödinger operator whose eigenvalues determine the hadron masses \cite{Boschi-Filho:2002xih, Erlich:2005qh}. A well-known example is the \emph{soft-wall} model, in which a static background dilaton breaks conformal invariance and introduces confinement \cite{Karch:2006pv, Colangelo:2008us, Gutsche:2012ez}. The dilaton defines a holographic potential in the Schrödinger-like equation satisfied by the bulk mode.

Isospectrality in AdS/QCD has been studied for scalar mesons \cite{Vega:2016gip}, and for vector mesons \cite{MartinContreras:2023eft}. Here, we focus on vector mesons. Mesons are unstable particles, and their decay properties play a central role in their identification. Considerable effort has been devoted to modeling decay constants of vector mesons within AdS/QCD \cite{Grigoryan:2010pj, Braga:2015jca, Braga:2017bml, MartinContreras:2021bis}.  

A notable finding of Ref. \cite{MartinContreras:2023eft} is that the isospectral transformation in AdS/QCD modifies the ground-state decay constant exclusively. Two dilatons connected by an isospectral transformation yield identical meson spectra except for the ground-state decay constant. This unique property motivates an investigation of the behavior of families of holographic vector mesons linked by an isospectral transformation when placed in a thermal medium (finite temperature or chemical potential). At zero temperature, the mass spectrum corresponds to the pole distribution of the retarded Green's function. In a thermal medium, the poles broaden into local peaks in the spectral functions, whose shapes evolve with temperature and chemical potential. This setting is ideal for studying the thermodynamics of the entire tower of radial excitations and, in particular, for clarifying the role of the ground-state decay constant in the complete dissociation of the quark-antiquark pair.

In this manuscript, we compute the spectral functions of the $\rho$ meson at finite temperature and chemical potential. First, we explore various holographic models within the same isospectral family and examine how the ground-state decay constant influences the thermal behavior of radial excitations. We find a clear monotonic relation between the ground-state decay constant of the $\rho$ meson and its melting temperature. Regarding thermal masses (defined as the peak positions in the spectral functions), different models within the same isospectral family exhibit distinct thermal masses. If one views isospectrality as an analog of a symmetry, the thermal mass signals its breaking at finite temperature or chemical potential. For excited states, we also observe small variations in the thermal masses across the isospectral family; the higher the radial excitation $n$, the smaller the deviations. 

We then fix the isospectral parameter to reproduce precisely the experimental ground-state decay constant of the $\rho$ meson and analyze the spectral functions of this specific model. The original soft-wall model fails to capture the ground-state decay constant and consequently underestimates the dissociation temperature. The isospectral transformation raises the decay constant and, as we show, raises the melting temperature accordingly. Our results agree well with other approaches and with available experimental data. For instance, we find a small suppression of the $\rho$ meson mass at finite temperature and a critical temperature for $\rho$ meson dissociation of $157~ \textrm{MeV}$. 

The article is organized as follows. In Sec. \ref{sec:2} we review the soft-wall holographic model for vector meson masses and decay constants. Section \ref{sec:3} discusses the isospectral transformation in bottom-up holography, which provides the starting point for the thermal analysis. Section \ref{sec:4} describes the methodology used to obtain spectral functions for the isospectral models at finite temperature and chemical potential. In Sec. \ref{sec:5} we present our results for the $\rho$ meson ground state, including the use of the isospectral parameter to obtain a functional dependence of the melting temperature on the decay constant, the effect of the isospectral parameter on the thermal mass, and the improvement of the holographic description of the $\rho$ meson in a thermal medium. Excited states are discussed in Sec. \ref{sec:6}. We conclude in Sec. \ref{sec:7}.

\section{Vector mesons in Bottom-up A\lowercase{d}S/QCD at zero temperature}
\label{sec:2}
As a starting point for considering the holographic approach to vector mesons, we have to set the geometrical background frame:

\begin{equation}\label{metric}
    dS^2=\frac{R^2}{z^2}\left[dz^2+\eta_{\mu\nu}\,dx^\mu\,dx^\nu\right]=\frac{R^2}{z^2}\,\eta_{MN}\,dx^M\,dx^N,
\end{equation}

\noindent where $R$ is the AdS radius. Latin indices run over five-dimensional quantities, and Greek indices over four-dimensional objects.  

Within the AdS/CFT correspondence, the identity of a hadron at the boundary is dictated by the conformal dimension $\Delta$ of its dual boundary operator. For vector mesons in the $s$-wave, this operator assumes the form $\mathcal{O}_\mu = \bar{q} \gamma_\mu q$, possessing a conformal dimension of $\Delta=3$. This dimension determines the mass of the dual bulk field propagating in the AdS space through the relation $M_5^2 R^2 = (\Delta - 1)(\Delta - 3)$, which yields $M_5^2 R^2 = 0$ for the vector case. 

The dynamics of the dual bulk vector field $A_M(x^\mu,z)$ is governed by a five-dimensional action, incorporating a static dilaton field $\Phi(z)=\kappa^2\,z^2$ to model confinement \cite{Karch:2006pv}:

\begin{equation}
    I = -\frac{1}{4\,g_5^2} \int d^5x \sqrt{-g} \, e^{-\kappa^2\,z^2}\, F_{MN}F^{MN},
\end{equation}

\noindent where $g_5^2$ is a normalization constant that fixes electromagnetic decay constants units, and $F_{MN}=2\partial_{\left[M\,\right.}A(x^\mu,z)_{\left.N\right]}$ is an abelian field strength. Regarding the quadratic profile dilaton, $\kappa$ sets the energy scale for vector meson masses. 

It is worth mentioning that this static quadratic dilaton approach works for light unflavored hadrons, described by linear radial Regge trajectories, $M_n^2=a(n+b)$, where $a$ is an \emph{almost} universal slope in energy squared units, and $b$ is an intercept that accounts for hadron spin $J$ or orbital angular momentum $L$. When this sort of model is extended to heavy mesons, such \emph{linearity} ceases, as Bethe-Salpeter analysis suggested \cite{Chen:2018bbr, Chen:2021kfw, Chen:2023djq}. This hypothesis leads to a different approach to static deform dilatons $\Phi(z)=\left(\kappa\,z\right)^{2-\alpha}$ consistent with non-linear Regge trajectories parametrized as $M_n^2=a(n+b)^\nu$ \cite{MartinContreras:2020cyg}. This idea was extended to melting temperature and thermal mass analysis for heavy quarkonia \cite{MartinContreras:2021bis}, hadronic stability \cite{MartinContreras:2024gsk}, $\Sigma$ baryons spectroscopy \cite{Guo:2024nrf}, and WKB analysis for heavy hadron decay modes \cite{Diles:2025xot}, and multiquark states spectroscopy \cite{MartinContreras:2025mpj}.   

The spectral properties of the boundary theory are encoded in the equation of motion for bulk fields. Upon implementing the transverse gauge $A_z(z)=0$, and performing a Fourier transform, the linearized equation of motion for the field $A_M(x^\mu,z)$ reduces to a Sturm-Liouville form. A subsequent Bogoliubov transformation 

\begin{eqnarray}
    \psi(z)&=&e^{\frac{1}{2}\,B(z)}\,u(z).\\
    B(z)&=&\
    \kappa^2\,z^2-\ln\left(\frac{R}{z}\right),
\end{eqnarray}

\noindent recasts this equation into a \emph{Schrödinger-like} equation for the hadronic modes as:

\begin{equation}\label{modes}
    -u''(z) + V(z) u(z) = M_n^2 u(z).
\end{equation}

The dilaton profile uniquely specifies the confining holographic potential $V(z)$ and, for vector mesons, is given by:

\begin{equation}\label{potential}
    V(z) = \frac{3}{4\,z^2} + \kappa^2\,z^2.
\end{equation}

The eigenvalues $M_n^2$ of this potential correspond to the squared masses of the vector meson radial trajectory, which could be identified with $\rho$ or $\omega$ mesons, since they have $J^{PC}=1^{--}$:

\begin{eqnarray}
    M_n^2=4\,\kappa^2\,n,\,\,n\in\mathbb{Z}^+.
\end{eqnarray}

The original softwall model is not sensitive to isospin information. For simplicity, we will choose the $\rho(770)$ meson trajectory. 

Beyond the mass spectrum, a crucial observable is the decay constant $f_n$, which quantifies the probability amplitude for a hadron to annihilate into the vacuum. Holographically, this is computed from the asymptotic behavior of the normalizable mode solutions $u_n(z)$ near the conformal boundary \cite{MartinContreras:2019kah}:

\begin{equation}
    f_n^2 = \frac{4}{M_n^2\,g^2_5} \lim_{z \to 0} e^{-2\kappa^2\,z^2} \left( \frac{R}{z} \right)^2 \left| \frac{u_n(z)}{z^2} \right|^2.
\end{equation}

The dilaton field $\Phi(z)=\kappa^2\,z^2$ plays the fundamental role of ensuring confinement, which is manifested in the potential $V(z)$ and, consequently, in the emergence of a discrete spectrum of normalizable modes $u_n(z)$ dual to the stable vector meson states. This established framework provides the foundation for subsequent analysis, such as applying isospectral transformations to study deformations of the quadratic dilaton $\Phi(z)$ and their implications for the thermal behavior of the spectral density.
\section{The isospectral paradigm in bottom-up A\lowercase{d}S/QCD}
\label{sec:3}

Isospectrality concerns the investigation of distinct mathematical objects, such as manifolds, differential operators, or quantum-mechanical potentials, that possess identical eigenvalue spectra \cite{abraham1980changes, darboux1882proposition, mielnik1984factorization, Bazeia:2002xg}. In non-relativistic quantum mechanics, this corresponds to different Hamiltonian potentials that yield the same set of energy eigenvalues. The formalism of supersymmetric quantum mechanics furnishes a powerful and systematic method for generating such families of isospectral potentials \cite{Witten:1981nf}, which generalizes the classical Darboux transformation. The procedure is initiated from a known potential $V_1(z)$, its associated ground state eigenfunction $\phi_0(z)$, and the corresponding spectral set $\{\lambda_n\}$ \cite{Cooper:1994eh}.

The mathematical construction is orchestrated via a superpotential $W(z)$, intrinsically linked to the ground state through the relation $W(z) = -d(\log \phi_0)/dz$, such that the original potential is factorized as $V_1(z) = W^2 - W'$. The central objective is to ascertain a family of superpotentials $\hat{W}(z)$ that reproduce the same superpartner potential $V_2(z) = W^2 + W'$. Imposing this \emph{isospectrality condition} leads to the definitive form of the new superpotential:

\begin{equation}
\hat{W}(z) = W(z) + \frac{d}{dz} \log\left[I(z) + \lambda\right],
\end{equation}

\noindent where $\lambda \in \mathbb{R}$ is an integration parameter and $I(z) = \int_0^z \phi_0^2(z')  dz'$ is a function defined from the ground state. The ensuing monoparametric family of potentials, strictly isospectral to $V_1(z)$, is then given by:

\begin{equation}
\hat{V}_\lambda(z) = \hat{W}^2 - \hat{W}' = V_1(z) - 2\frac{d^2}{dz^2} \log\left[I(z) + \lambda\right].
\end{equation}

Although the potential $\hat{V}_\lambda(z)$ shares the same wave-functions with $V_1(z)$ in its excited states, its ground state is distinct. The modified ground state wavefunction for the deformed potential is not obtained from a direct spectral solution. Still, it is provided analytically by the expression $\phi_{0,\lambda}(z) = \phi_0(z) / (I(z) + \lambda)$.

This formalism provides a robust analytical tool for deforming a given potential, enabling the construction of continuous isospectral families that preserve spectral data while altering the underlying potential structure, a feature with significant implications for fields such as holographic QCD \cite{Vega:2016gip, MartinContreras:2023eft}.

For a holographic model defining a hadron mass spectrum, there is a one-parameter class of equivalent models providing the same hadron masses. An interesting finding of previous works \cite{Vega:2016gip, MartinContreras:2023eft} is that the isospectral transformation changes the electromagnetic decay constant of the ground state, while keeping the decay constant of the excited states untouched. It allows us to start with a quadratic (or other choice) dilaton and use the isospectral transformation to adjust the ground-state decay constant. For the $\rho$ meson, we can fit masses using the quadratic dilaton $\Phi(z)=k^2 z^2$ and fine-tuning its ground state decay constant using the isospectral transformation \cite{MartinContreras:2023eft}. Finally, we reconstruct the dilaton associated with the fine-tuned decay. 

The picture we have for the quadratic dilaton is as follows. Considering the AdS description in terms of the metric eq.\eqref{metric}, the Regge linear spectrum $m_n^2\propto n$ leads to the quadratic dilaton $\Phi(z)=k^2z^2$ which in turn defines the holographic potential of eq.\eqref{potential}. The isospectral transformation then generates the class of potentials parameterized by $\lambda$:

\begin{equation}\label{isospectral}
    \hat{V}_\lambda(z)=V(z)
    -2\,\frac{d^2}{d\,z^2}\,\log \left[1-\Gamma\left(2,\kappa^2\,z^2\right)+
    \lambda\right],
\end{equation}

\noindent with $\Gamma(2,k^2z^2)$ the incomplete gamma function. Finally, we map the potentials $V_\lambda(z)$ into the family of isospectral dilatons $\Phi_\lambda(z)$ by a systematic reconstruction defined by 

\begin{equation}\label{dilaton-eng}
\hat{V}_\lambda(z)=\frac{3}{4\,z^2}+\frac{1}{2\,z}\Phi_\lambda^{\prime}(z)
+\frac{1}{4}\Phi_\lambda^{\prime}(z)^{2}-\frac{1}{2}\Phi_\lambda^{\prime\prime}(z)
.  
\end{equation}

\begin{figure}
  \includegraphics[width=3.4 in]{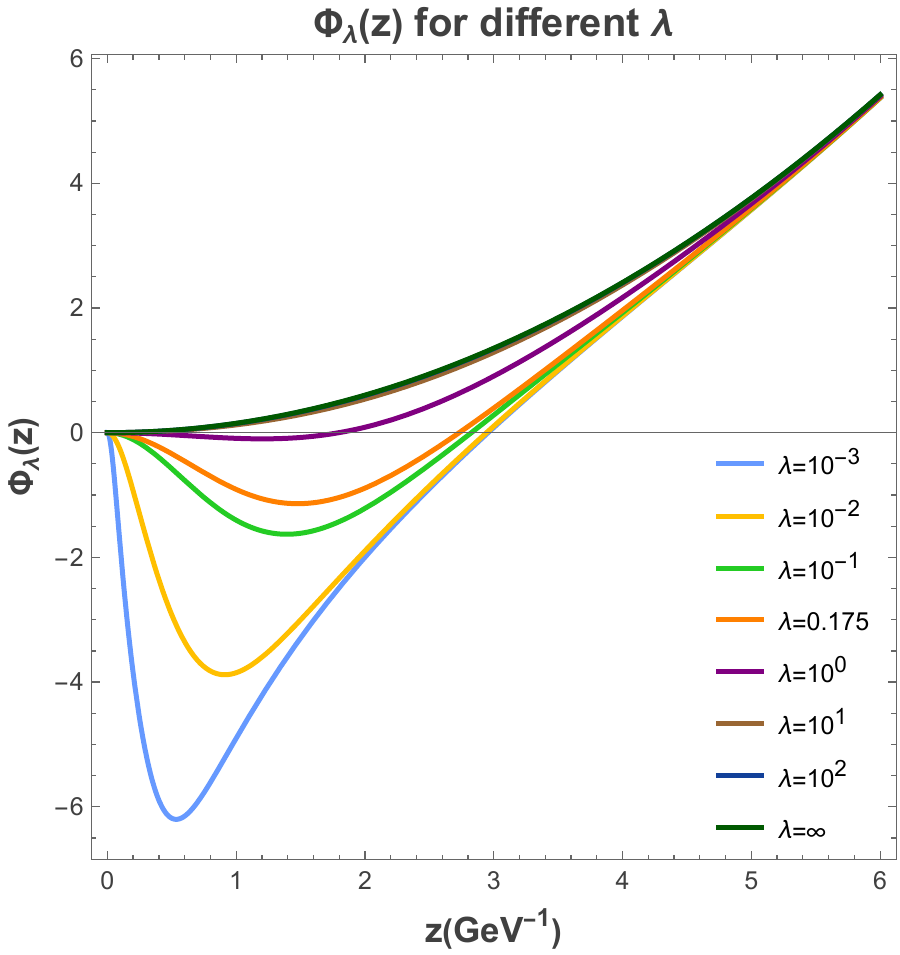}
\caption{Numerical reconstruction of the probe dilaton field $\Phi_\lambda(z)$ from the family of isospectral dilatons defined from the vector meson softwall model potential given in Eqn. \eqref{potential}. Notice that $\lambda\to\infty$ recovers the original softwall dilaton. The energy scale in this reconstruction is set by the $\rho(770)$ experimental mass \cite{ParticleDataGroup:2024cfk}. For computational purposes, we choose our \emph{numerical infinity} at $\lambda=9999$, where the deviation in comparison with the non-isospectral dilaton is negligible.}
\label{fig:onea}
\end{figure}

It is the reconstructed dilaton $\Phi_\lambda(z)$ that defines the bottom-up AdS/QCD model, which shares the same mass spectrum as the soft-wall model. We remark that the reconstructed dilatons associated with the family of isospectral potentials are non-analytic, even when we start from the quadratic dilaton. We present in Fig. \ref{fig:onea} some of the isospectral dilaton partners of the standard quadratic profile, $\Phi(z)=\kappa^2\,z^2$.

In summary, the isospectral parameter $\lambda$ defines the holographic model through dilaton reconstruction from the isospectral potential.

Once we have fixed one of the isospectral dilatons by setting $\lambda$, we use it as the new probe dilaton $\Phi_\lambda(z)$ and write it in the action for vector mesons as: 

\begin{equation}\label{vector-field-action}
    I_\lambda = -\frac{1}{4\,g_5^2} \int d^5x \sqrt{-g} \, e^{-\Phi_\lambda(z)}\, F_{MN}F^{MN}.
\end{equation}

Here $g_{MN}$ are the components of the AdS metric defined in eq.\eqref{metric}.

Then, we proceed to finite temperature and \emph{finite chemical potential} by introducing a black hole in the static bulk geometry. Therefore, the choice of the spectral parameter at zero temperature determines the background dilaton, which is then used to study thermal effects.  We compare the thermal evolution of different dilatons sharing the same mass spectrum and analyze the thermal-medium effects of the isospectral transformation by examining their spectral functions.  

\section{Finite temperature and finite density extensions: vector meson case}
\label{sec:4}

The calculation of the finite-temperature and finite density spectral function proceeds from the holographic prescription for the retarded correlator of the boundary electromagnetic current $J^\mu = \bar{Q} \gamma^\mu Q$, which is dual to the massless bulk field $A_M(q,z)$ \cite{Son:2002sd}, whose action follows the structure exposed in the equation \eqref{vector-field-action}. This structure is standard for vector mesons, and we can use it for light-unflavored vector mesons \cite{Fujita:2009ca, Fujita:2009wc}.

The recipe for the isospectrality inclusion is as follows: for a given $\lambda$, we compute the associated isospectral dilaton $\Phi_\lambda(z)$ numerically. Next, we plug this dilation into the action \eqref{vector-field-action} and extract the spectral density.

The geometry background is set to be a five-dimensional AdS-Black hole background defined by the line element:

\begin{eqnarray}
     dS^2&=&\frac{R^2}{z^2}\left[\frac{dz^2}{f(z)}-f(z)\,dt^2+d\vec{x}\cdot\,d\vec{x}\right],\\
     f(z)&=&\left(1-\frac{z}{z_h}\right)\,G(z)
\end{eqnarray}

\noindent where $z_h$ is the locus of the most external event horizon. The function  $G(z)$ depends on other black hole quantities such as chemical potential \emph{via} the $U(1)$ Abelian charge in the Reissner-Nordstrom solution or the angular momentum in the rotating case.  The Hawking temperature in this case takes the form: 

\begin{equation}
    T=\frac{\left|G(z_h)\right|}{4\,\pi\,z_h}.
\end{equation}

We will consider thermal and finite density effects. Thus, we fix the blackening factor to be of AdS-Reissner-Nordstrom type:

\begin{eqnarray}
    G(z)&=&\left(1+\frac{z}{z_h}\right)\left(1+\frac{z^2}{z_h^2}-Q^2\,\frac{z^4}{z_h^4}\right),\\
    T&=&\frac{1}{\pi\,z_h}\left(1-\frac{Q^2}{2}\right),\\
    \mu&=&\frac{Q}{z_h}.
\end{eqnarray}

Notice that $R$ is the AdS radius and $g_5=2\,\pi$ fixes bulk action units, and also fixes the decay constant units \cite{Erlich:2005qh}. 

From the bulk action and the metric, it is possible to write the \emph{Sturm-Liouville} form of the equations of motion, in Fourier space, as 

\begin{eqnarray}\label{thermalmode1}
\partial_z\left[e^{-B_\lambda}\,\partial_z\,A_t\right]+\frac{e^{-B_\lambda}}{f}\left[\frac{\omega^2}{f}-k^2\right]\,A_t&=&0\\\label{thermalmode2}
\partial_z\left[f\,e^{-B_\lambda}\,\partial_z\,A_i\right]+e^{-B_\lambda}\,\left[\frac{\omega^2}{f}-k^2\right]\,A_i&=&0
\end{eqnarray}

\noindent where $A_t(z,q)$ stands for the vector field time component, $A_i(z,q)$ for the spatial one and $q_\mu=(\omega,k_i)$ is the boundary wave vector. It is worth mentioning that we have imposed the transverse gauge $A_z(z,q)=0$, which is consistent with the Ward identity $i\,q_\mu\, A^\mu = 0$.  The function $B$ present in the equations above is defined as 
\begin{equation}
    B_\lambda(z) = \Phi_\lambda(z) -\textrm{log}\left(\frac{R}{z}\right).
\end{equation}

Since we want to compute the thermal retarded Green's function, we will focus on the spatial part. We will also consider the rest case, where $k_i=0$ so that $q=\omega$. To compute the Green's function, we require the \emph{bulk-to-boundary propagator}, $\mathcal{V}(z,\omega)$ introduced as $A_i(z,\omega)=\tilde{A}_i(\omega)\,\mathcal{V}(z,\omega)$, which is normalized as $\mathcal{V}(0,\omega)=1$. For numerical simplicity, we will perform the transformation $u=z/z_h$, which sets the horizon to $u=1$. Therefore, the bulk-to-boundary equation is written as

\begin{equation}\label{bulk-to-boundary}
    \partial_u\left[e^{-B_\lambda(u)}\,f(u)\,\partial_u\,\mathcal{V}(z,\omega)\right]+\frac{z_h^2\,\omega^2}{f(u)}\,e^{-B_\lambda(u)}\,\mathcal{V}(z,\omega)=0.
\end{equation}

The \emph{retarded Green function}, whose imaginary part defines the spectral density, is constructed following the \emph{Minkowskian prescription} \cite{Son:2002sd} as

\begin{equation}\label{Green-Th}
    G_R(\omega,\lambda) = -\frac{2\,R}{z_h\,\mathcal{N}} \left. e^{-B_\lambda(u)} f(u) V(u,-\omega) \partial_u V(u,\omega) \right|_{u=0},
\end{equation}
\noindent where $\mathcal{N}=2\,R/g_5^2$ is a normalization constant that arises from the overall normalization of the bulk action and the holographic dictionary for the retarded Green's function. It combines $g_5$ and $R$, and it is fixed when we require the proper ultraviolet behavior for the holographic two-point consistent with the operator product expansion at large $N_c$ \cite{Erlich:2005qh}. In practice, 
$\mathcal{N}$ cancels out when we consider ratios of spectral functions, or when we match the numerical results to asymptotic conditions (as in data-driven approaches \cite{Chen:2024mmd}). Thus,  its precise value does not affect the extracted physical quantities such as peak positions, widths, or melting temperatures.

A general solution for the equation \eqref{bulk-to-boundary} can be written as a linear combination of two independent \emph{horizon-regular} solutions $\psi_0(u,\omega)$ and $\psi_1(u,\omega)$ in such way as \cite{MartinContreras:2021bis}

\begin{equation}
    \mathcal{V}(u,\omega)=\left[\psi_1(u,\omega)+\frac{B}{A}\psi_0(u,\omega)\right],
\end{equation}

\noindent where $A$ and $B$ are constants, and  $\psi_{0,1}(u,\omega)$  are  functions chosen so that near the conformal boundary they reduce to the asymptotic forms

\begin{eqnarray}
    \psi_0(u,\,\omega) &=& \frac{2}{\omega \,z_h}u\,J_1(\omega\, z_h\, u),\\
\psi_1(u,\,\omega) &=& -\frac{\pi\,\omega\, z_h}{2}u\,Y_1(\omega\, z_h\, u).
\end{eqnarray}

On the horizon side, we required that $\mathcal{V}(u,\omega)$ meets the \emph{out-going boundary condition} given by

\begin{equation}
    \mathcal{V}(u\to1,\omega)=\left(1-u\right)^{-\frac{i\,\omega\,z_h}{4}}.
\end{equation}

With all of these expressions, we can write the spectral density as

\begin{equation}
    \rho_\lambda(\omega)=-\operatorname{Im}\,G_R(\omega,\,\lambda)=\frac{2\,R}{z_h\,\mathcal{N}}\,\operatorname{Im}\,\frac{B}{A}.
\end{equation}

This expression is evaluated numerically by fixing $\lambda$ and running for selected values of temperature $T$ and chemical potential $\mu$.  
\begin{center}
  \begin{figure*}
    \centering
    \includegraphics[width=1.7 in]{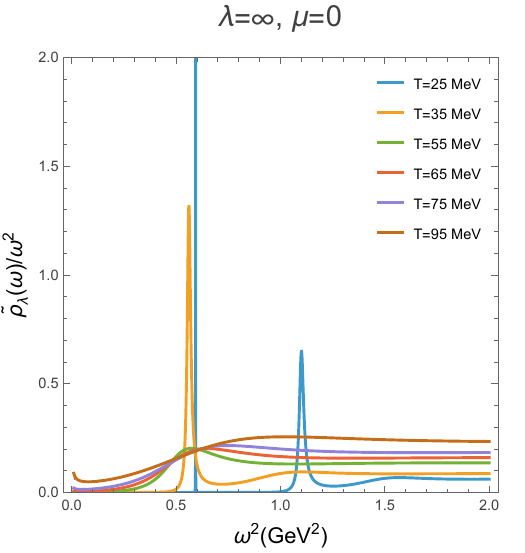}
    \includegraphics[width=1.7 in]{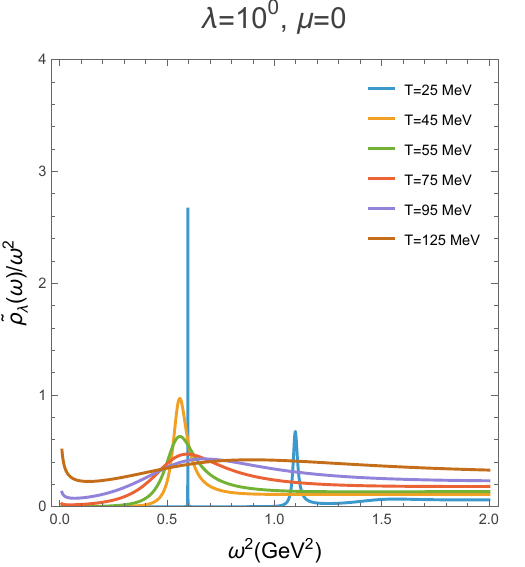}
    \includegraphics[width=1.7 in]{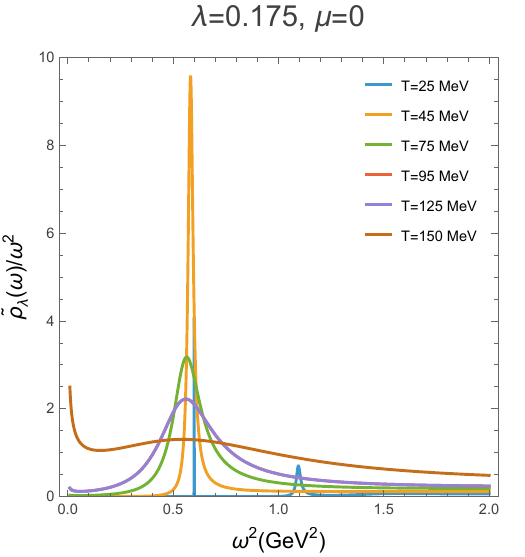}
    \includegraphics[width=1.7 in]{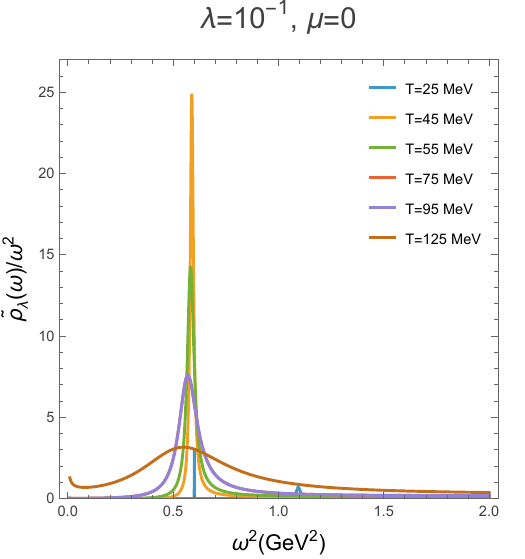}
    \caption{Melting process for different values of $\lambda$. Each panel fixes the chemical potential $\mu=0$ and runs over temperature $T$.}
    \label{fig:a}
\end{figure*}  
\end{center}

\begin{figure*}
 
  \includegraphics[width=7.0 in]{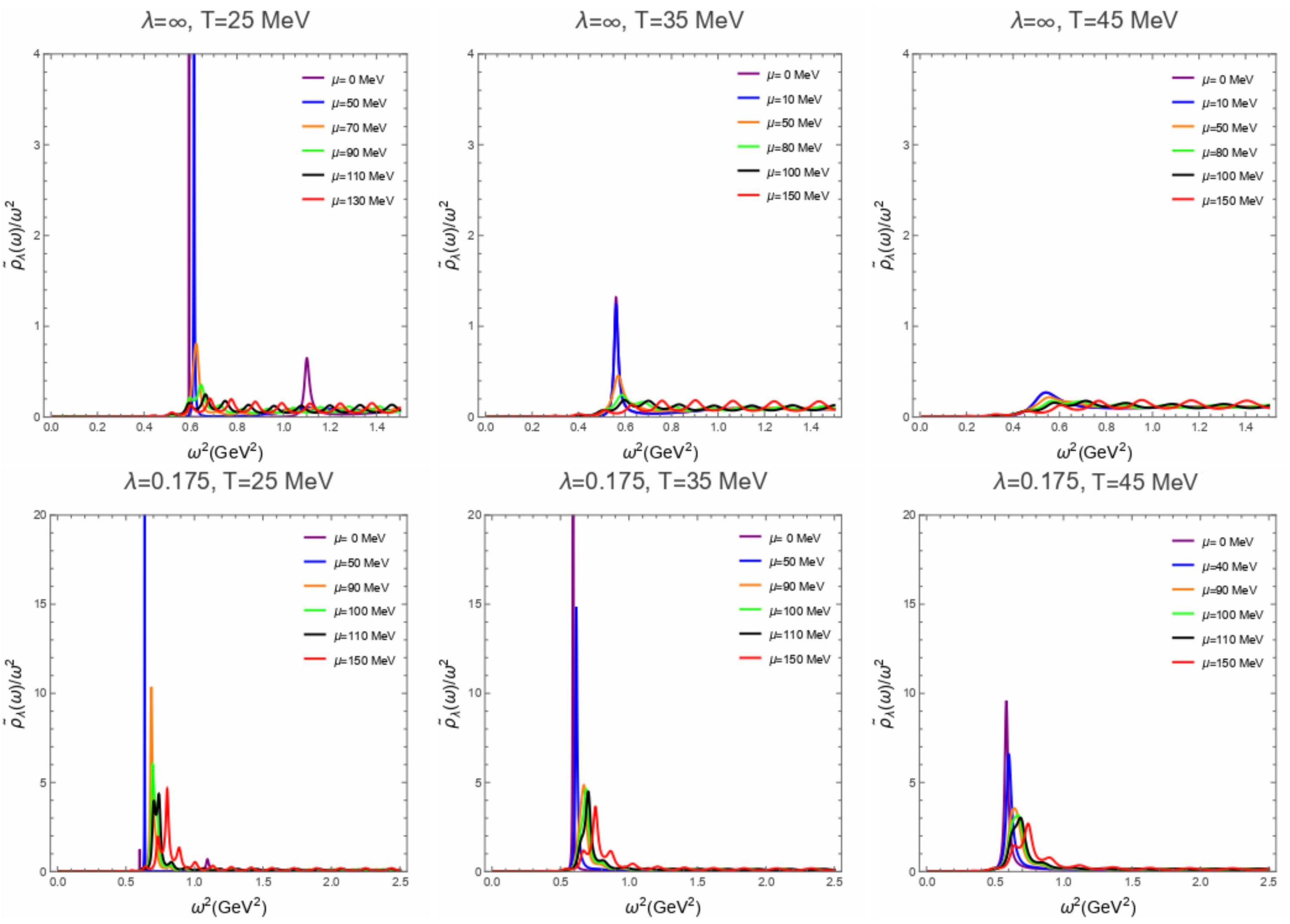}

 \caption{Melting process for different values of $\lambda$. Each panel fixes the temperature $T$ and runs over the chemical potential $\mu$. We have defined $\tilde{\rho}_\lambda(\omega)\equiv\mathcal{N}\,\rho(\omega)/2\,R$}
 \label{fig:b}
\end{figure*}

\begin{center}
  \begin{figure*}
    \centering
    \includegraphics[width=2.3 in]{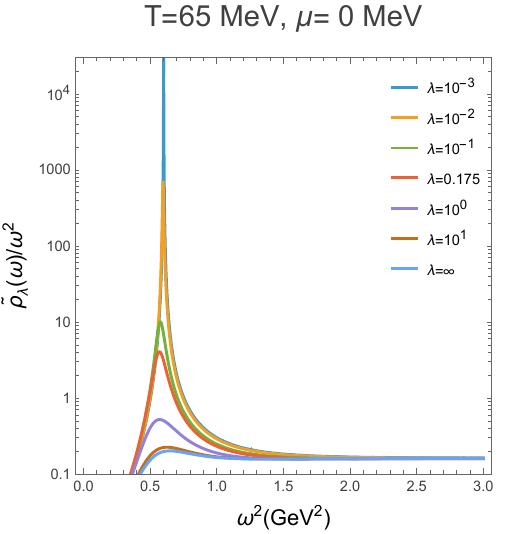}
    \includegraphics[width=2.3 in]{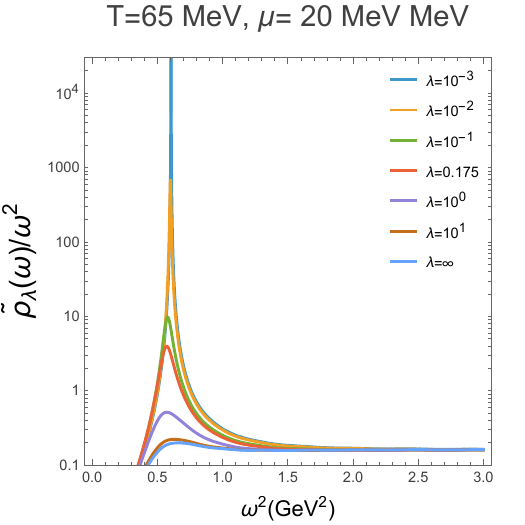}
    \includegraphics[width=2.3 in]{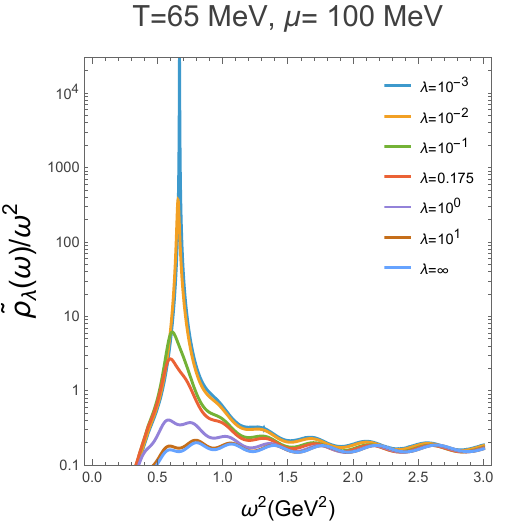}\\
    \includegraphics[width=2.3 in]{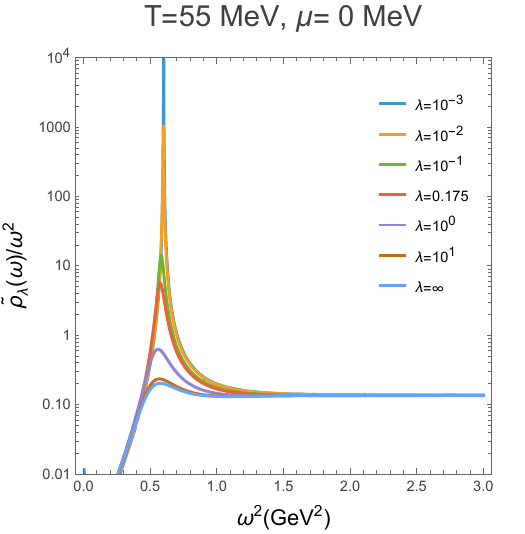}
    \includegraphics[width=2.3 in]{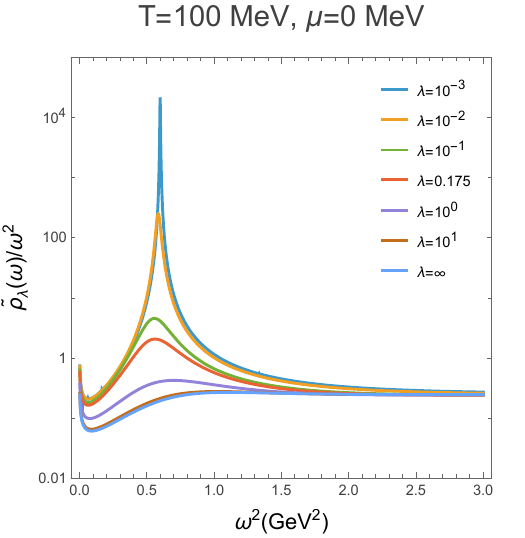}
    \includegraphics[width=2.3 in]{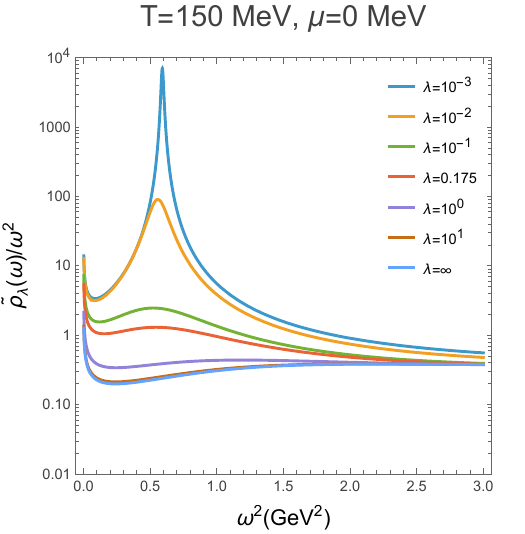}\\
    \caption{Melting picture considering $\lambda$ running. The upper panel shows the progression when the temperature is fixed while the chemical potential varies. The lower panel considers the opposite: fix the chemical potential and vary the temperature.}
    \label{fig:c}
\end{figure*}  
\end{center}

\section{Dissociation temperature, decay constant, and the isospectral parameter}
\label{sec:5}
\subsection{Melting Temperature versus decay constants}
For neutral vector mesons, like $\rho(770)$, the electromagnetic decay constant $f_n$ carries information about how compact the local quark density is: it is a measure of how much of the hadron wave function is concentrated at the origin, i.e., \emph{Van Royen-Weisskopf approach}. It also tells how likely the $q\bar{q}$ pair will be annihilated into virtual photons: a diffuse meson wavefunction implies that quarks are rarely at the same spatial point, leading to weak electromagnetic coupling. In other words, the value of $f_n$ for a given vector meson $V$ in the $V\to e^+\,e^-$ process dictates how strongly it is coupled to the electromagnetic current and hence, how compact the meson $V$  is. This fact is directly visible in the $e^+\,e^-$ cross-section: the peak height (associated with the integrated area) scales with $f_n^2$. \par

The latter sentence leads to the spectral interpretation of the electromagnetic decay constant: $f_n^2$ gives the \emph{probability} that the $e^+\,e^-$ current creates a meson from the vacuum \cite{Shifman:1978by}. It is precisely in the sum rule context where, in the zero-width limit, $f_n$ defines the residue of the spectral density: $\rho(q^2)\propto f_n^2\,m_V^2\,\delta\left(q^2-m_V^2\right)$. In a quantum mechanical sense, $f_n$ is the creation amplitude for the meson from the vacuum by the current, analogously to the pion decay constant, which gives the amplitude for the axial current to create a pion \cite{Shifman:1978bx}.\par

When we consider finite-density or finite-temperature cases, the colored medium causes the hadron to lose its identity as a bound state. This process is known as \emph{melting}.  The information of how the medium affects a given vector meson is encoded in the spectral density $\rho(\omega)$. In this context, the decay constant is promoted to be a function of temperature and baryonic chemical potential, i.e., $f_n(T,\mu)$, and it still has information about whether the quarks are still correlated into a compact meson. If $f_n(T,\mu)$ remains sizable, the quark-antiquark pair still forms a well-defined composite object. If $f_n(T,\mu)\to0$, the bound-state is undergoing a \emph{melting process}, since the quark-antiquark correlation is ceasing. Thus, the decay constant $f_n(T,\mu)$ \emph{behaves similarly} to an \emph{order parameter} for the melting process \cite{Ayala:2012ch}.\par

From the spectral density point of view, the meson acquires a width from collisions with the heat bath/dense medium. This is translated into the delta pole broadening into a quasiparticle finite-height peak, modeled by a Breit-Wigner distribution \cite{Breit:1936zzb}. In the case of light mesons, the evolution of $f_n(T,\mu)$ deals with the chiral symmetry restoration. In the heavy quarkonia sector, the suppression of $f_n(T,\mu)$ is a signal of Debye screening. Thus, in the hot and dense medium, the decay constant becomes a direct \emph{agent} for the survival of the meson as a well-defined state. Its vanishing is the unambiguous signal of the melting process.

At the hot and dense medium, the decay constant $f_n(T,\mu)$ is still the residue of the complex pole, or equivalently, the quantity $f_n(T,\mu)^2\, M_n^2(T,\mu)$ is proportional to the area under the quasiparticle peak at $M_n(T,\mu)^2$. Thus, the peak vanishing in the spectral density provides a phenomenological test for assessing whether melting is occurring \cite{MartinContreras:2021bis}. 

In the holographic context, the modeling of electromagnetic decay constants $f_n$ for vector mesons in bottom-up models has been addressed in \cite{Grigoryan:2010pj, Braga:2015jca, Braga:2015jca, MartinContreras:2019kah, MartinContreras:2021bis}. All of these references have pointed out that for dilaton-based models with the condition $\Phi(z\to0)=0$ lead to ill-behaved decay constants, i.e., they do not decrease with the radial excitation number $n$. Furthermore, in addition to the ill-behaved $f_n$ spectrum, the softwall model has another issue: the melting temperatures for $\rho$ mesons, read from the spectral density, seem to be lower than their phenomenological counterpart. Since decay constants at zero temperature and melting temperatures (where the quasiparticle peak vanishes) are connected, it is natural to say that having ill-behaved decay constants is reflected in having lower melting temperatures.  In the pure bottom-up holographic context, this issue translates into how the lowest coefficient $c_1$ in the expansion of bulk modes is constructed, \emph{i.e.}, 

\begin{equation*}
   \left. \psi(z)\right|_{z\to0}=c_1\,z^{\Delta+1}+\mathcal{O}\left(z^m\right)
\end{equation*}
\noindent with $m>\Delta+1$, since $c_1$ controls the holographic decay constants behavior \cite{MartinContreras:2019kah}.

In contrast, the deconfinement phase transition temperature for the softwall model, computed from the Hawking/Page transition, seems to agree with phenomenological expectations (See Table \ref{tab:temps}). This situation arises from the decay constants, which are not necessary for computing the Hawking/Page phase transition. This discrepancy between the meson melting and the Hawking/Page phase transition temperature is a drawback of the dilaton-based models with $\Phi(z\to0)=0$, as cited in \cite{Vega:2017dbt, MartinContreras:2021bis}.

However, new proposals, such as isospectrality \cite{MartinContreras:2023eft} and data-driven models \cite{Zhang:2026zoz}, seem to shed light on this discussion from different perspectives. In the latter case, a dataset consisting of heavy vector meson masses and decay constants is used to train a perceptron that yields a holographic potential and dilaton that reproduce the expected phenomenological behavior. In the former case, authors analyze four different bottom-up dilaton-based models. The isospectral transformation modified the ground-state decay constant. It does so by deforming the holographic potential by producing a well near the boundary at $z=0$ in such a specific form that the decay constants of the excited states remain unchanged. 

For readiness purposes, we consider the isospectral extension of the softwall model. Recall that decay constants in this approach, for vector mesons, are degenerate. When isospectral transformations, as depicted in the expression \eqref{isospectral}, are applied, this degeneracy in the ground state is lifted. Thus, in the case of light vector mesons, as $\rho(770)$, we can fine-tune $\lambda$ to match the expected decay constant value, $f_1^{\rho(770)}=221.21\pm0.21$ MeV \cite{ParticleDataGroup:2024cfk}. Thus, the main hypothesis to test in this framework is that $\lambda$ increases the $\rho(770)$ melting temperature to meet phenomenological and experimental data.

The methodology we will follow to test this hypothesis is as follows: we apply the methodology discussed in the previous section to examine the effect of the isospectral transformation on the properties of the hot or dense medium. At zero-chemical potential, we have that for each value of $\lambda$ there is one ground state decay constant $f_1(\lambda)$ and one melting temperature $T_m(\lambda)$ and we compose functions to obtain $T_m=T_m(f_1)$, which is the melting temperature as a function of the ground state decay constant. 

We define the melting temperature $T_{\mathrm{melting}}$ as the temperature at which the ratio of the peak height to its width (full width at half maximum) falls below unity, indicating that the quasiparticle resonance is no longer distinguishable from the continuum.

\begin{figure*}
 
  \includegraphics[width=4.6 in]{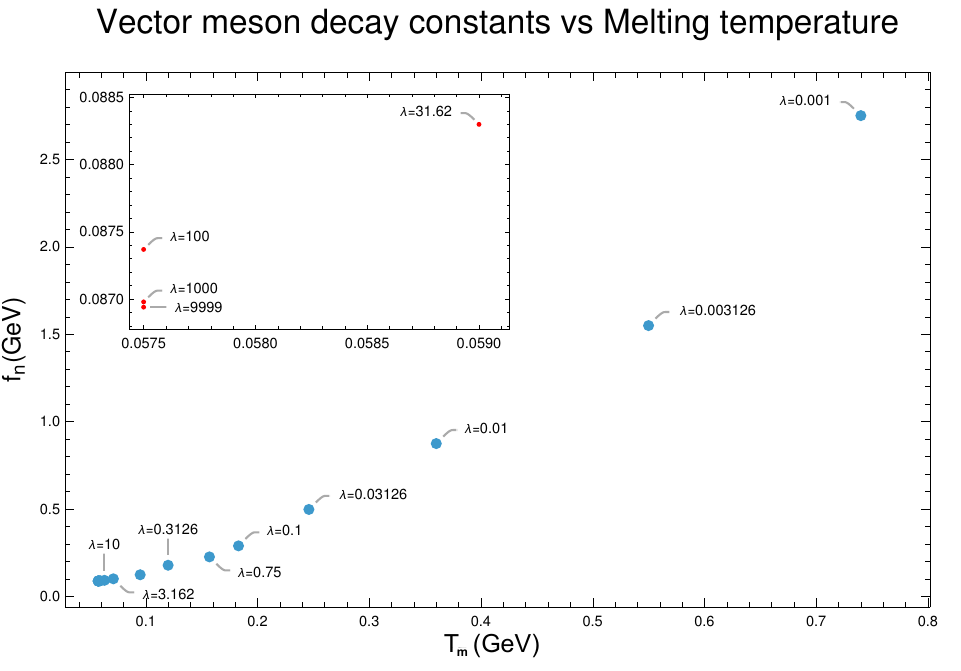}
\caption{Electromagnetic decay constant for vector meson ground state as a function of the melting temperature. Notice that $ T_m$ is monotonic increasing with $f_1$.}
\label{fig:melting}
\end{figure*}

 We show in Figure \ref{fig:melting} our results for the $\rho$ meson melting temperature as a function of its ground state decay constant at zero temperature. We observe that the greater the ground state decay constant, the higher the temperature for the $\rho$ meson melting. The growth is not linear, but it is clearly monotonic. This observation supports the interpretation of the decay constant as an energy scale that controls meson dissociation in the thermal medium. We remark that the decay constant is not a binding energy. Nevertheless, semiclassical expressions such as the \emph{Segre-Fermi formula} allow us to connect the spectrum of decay constants to the mass spectrum \cite{Diles:2025xot}. This implies that, at least at the semiclassical level, the decay constant encodes information about the hadronic binding energy, reinforcing the observed correlation between $f_1$ and $T_m$.

Isospectrality at finite temperature was previously discussed in \cite{Afonin:2018era}, where the isospectral parameter $\lambda$ is shown to change the critical temperature of the Hawking-Page transition, which is interpreted as the deconfinement transition in the dual gauge theory \cite{Hawking:1982dh, Herzog:2006ra, Braga:2024nnj}. Curiously, in \cite{Afonin:2018era}, it is found that the critical temperature for the Hawking-Page transition increases with $\lambda$. In the particular case of the softwall model, deconfinement temperature \cite{Herzog:2006ra} is reached when $\lambda=100$. Recall that the original potential and its isospectral partner become equal when $\lambda\to\infty$. Thus, the observed effect in the isospectral deconfinement temperature is due to the \emph{isospectral reconstruction} of the standard softwall confinement potential \cite{Karch:2006pv}.\par

In the spectral function case, numerical experiments demonstrate that increasing $\lambda$ decreases the melting temperature (see Fig. \ref{fig:c}), even below the corresponding Hawking/Page temperature. This behavior is not surprising despite how the zero temperature decay constants vary with $\lambda$: since decay constants decrease when $\lambda$ increases, the pole residue at $M_\rho$ should also decrease, implying the same for the $\rho(770)$ meson melting temperature. Fig. \ref{fig:melting} summarizes this scenario. 

On the one hand, it is not surprising that the results disagree in their dependence on $\lambda$, even though, in both cases, we obtain a critical temperature associated with the same phenomenon of quark-gluon plasma formation \cite{Busza:2018rrf}. The reason for this disagreement is \emph{methodological}. In one approach, the Hawking-Page transition in a static background with an isospectral dilaton probe is considered, whereas the other considers the disappearance of the quasiparticle peak. In the former case, the isospectral probe-dilaton affects the free energy differently from the way the same object affects the bulk hadron states. At the thermodynamic level, inserting a probe static dilaton produces a background that is not a saddle point of the full action. See, for example, the Einstein-Maxwell-Dilaton model \cite{Li:2014dsa}.
 
For the spectral density case, the analytical behavior of the deformed $\Phi_\lambda(z)$ is such that for smaller $\lambda$ the dilaton presents a deeper well near $z=0$, which produces bound-states, implying large melting temperatures for the spectral peaks. This scenario favors the Black-Hole-AdS solution over the Thermal AdS one. A similar situation happens when we consider chemical potential.

The effect of a finite baryon chemical potential $\mu$ on the $\rho$ meson spectral function is illustrated in the upper panels of Fig. \ref{fig:c}. At a fixed temperature $T$, increasing $\mu$ progressively suppresses the quasiparticle peak, eventually driving the meson into the continuum. This behavior is expected: both thermal and density effects weaken the $q\bar{q}$ binding and lead to dissociation, in agreement with general analyses of holographic spectral functions at finite baryon density \cite{Mas:2008jz, Georgiou:2022ekc}. A more refined picture emerges when we examine the dependence on the isospectral parameter $\lambda$ while holding $T$ and $\mu$ fixed. For any given $(T,\mu)$, the peak height decreases monotonically with $\lambda$. Since a larger $\lambda$ corresponds to a smaller zero‑temperature decay constant $f_1$ (see Fig. \ref{fig:melting}), this observation directly confirms that $\lambda$ fine‑tunes the ground‑state electromagnetic decay constant, which acts as the spectral density residue at the pole mass, even in a hot and dense medium. The isospectral parameter thus controls the survival of the $\rho$ meson across the entire $(T,\mu)$ plane. This interplay among temperature, chemical potential, and the dilaton deformation is reminiscent of the Einstein‑Maxwell‑Dilaton (EMD) approach to holographic QCD, in which the phase diagram and the deconfinement transition are systematically studied \cite{Cai:2012xh}. The present isospectral construction complements those analyses by providing a clean separation between mass spectrum and decay constant, enabling a precise study of how the latter influences melting in a dense environment.

\begin{table*}[t]
    \centering
    \begin{tabular}{c|c|c}
         \hline
         \multicolumn{3}{c}{\textbf{Deconfinement Temperature}}\\
         \hline
         \textbf{Source} & $T_c$(MeV) & \textbf{Method}\\
         \hline
         \hline
        ALICE \cite{Baral:2026ohe,Flor:2021olm} & $155-165$ & Thermal-FIST  (chemical freeze-out)\\
        HotQCD\cite{HotQCD:2018pds}& $156.5\pm1.5$  & HISQ  \\
        Gavai et al. \cite{Gavai:2024mcj} &$158.7^{+2.6}_{-2.3}$ &(2+1) flavor QCD with M\"obius domain-wall\\
        Mickley et al. \cite{Mickley:2024vkm} & $321\pm6$& Center vortex analysis   \\
        Herzog \cite{Herzog:2006ra} & $191$ & Softwall Model for $\rho$ meson\\
        Herzog \cite{Herzog:2006ra} & $125$ & Hardwall Model for $\rho$ meson\\
        Kim et al. \cite{Kim:2007em} & $100-130$ & Hardwall for $\rho$ meson with quark backreaction\\
        Afonin et al. \cite{Afonin:2014jha} & $260$ & Softwall with planar gluodynamics\\
        Afonin \cite{Afonin:2018era} & $235-305$ & Isospectral Softwall for scalar glueball $f_0(900)$\\
        Chen \cite{Chen:2024mmd} & $147$ & EMD-Data driven $2+1$ model\\

       Martin et al. \cite{MartinContreras:2025wnh}
         & $169$ & Reconstructed WKB.\\
         \multirow{2}{*}{\textbf{This work}}
         & $157$ & Isospectral softwall model (quasiparticle peak melting, $\lambda=0.175$).\\ 
        & $118.4$ & Deconfinement phase transition with isospectral dilaton, $\lambda=0.175$.
    \end{tabular}
    \caption{Summary of deconfinement temperatures for different analyses (Experimental, non-holographic, and holographic). }
    \label{tab:temps}
\end{table*}

\subsection{Thermal mass and the isospectrality}
In a hot or dense medium, the notion of an isolated particle is replaced by the quasiparticle, which appears as a broadened peak in the spectral function. The position of this peak defines the quasiparticle mass, and generally depends on temperature and/or baryon density. 

Earlier numerical experiments with our model showed that isospectral transformations can nevertheless strongly increase the decay constant for the vector meson ground state. Such an observation suggests that the spectral peak location alone does not fully characterize the medium response, prompting us to ask about the role isospectrality plays. 

At zero temperature,  the mass of a stable meson is also dictated by the current-current correlator pole. At finite temperature, this pole is replaced by a finite-width peak in the corresponding spectral function, allowing one to read the thermal mass from the peak position, see figure \ref{fig:0175}. This standard prescription has been widely adopted in AdS/QCD models with in-medium hadrons \cite{Colangelo:2009ra,cui2016finite}. \par

Here, we investigate whether the existence of a family of dilatons that define the same zero-temperature mass spectrum implies the same thermal masses, provided the corresponding states survive. We first remark that the result of the previous section should be taken into account here. Different values of the isospectral parameter yield distinct dissociation temperatures, indicating that the thermal mass temperature domain is sensitive to this parameter.\par

Notice that for a fixed temperature, varying $\lambda$ from $0$ to $\infty$ results in a slight shift in the quasiparticle peak location. Such a displacement corresponds to a small variation in the thermal mass: for a given temperature, the thermal ground‑state mass changes with $\lambda$. This indicates that isospectrality (defined for the zero‑temperature Schrödinger operator) does not extend trivially to the finite‑temperature spectral function. The observed variation is small compared with the vacuum hadron mass. The same sensitivity to $\lambda$ appears at finite chemical potential. In Figure~\ref{fig:c}, the temperature is high enough to suppress spectral peaks of the excited states; those are discussed in the next section.\par

The fact that thermal masses differ among isospectral partners is not unexpected. The isospectral transformation modifies the probe dilaton $\Phi_\lambda(z)$, which enters the action, and thus the thermal Green's function. Although the zero‑temperature mass spectrum is identical by construction, the thermal medium breaks the degeneracy because the bulk‑to‑boundary propagator depends on the detailed shape of $\Phi_\lambda(z)$ through the function $B_\lambda(z)$ in Eq.~\eqref{Green-Th}. In other words, the isospectral transformation is defined for the vacuum Schrödinger equation \eqref{modes}, not for the thermal wave equation \eqref{thermalmode2}. We therefore observe a small, expected violation of isospectrality in the thermal masses.

We emphasize that isospectrality guarantees identical eigenvalues of the zero‑temperature Schrödinger operator, not identical thermal Green's functions. The observed variation in thermal masses is therefore a consistency check, not a flaw.

\subsection{Hawking–Page transition and the isospectral dilaton}

We now examine the confinement–deconfinement transition \emph{via} the Hawking–Page analysis \cite{Herzog:2006ra}. In this approach, the deconfinement temperature $T_c$ is obtained by comparing the free energies of two backgrounds: thermal AdS (confined phase) and the AdS–Schwarzchild black hole (deconfined phase), with the isospectral dilaton $\Phi_\lambda(z)$ included as a probe. The Euclidean action for the background geometry takes the form:

\begin{equation}
    I_\text{b}=\frac{1}{8\,\pi\,G_5}\int d^5x\,\sqrt{-g_\text{b}}\;e^{-\Phi_\lambda(z)}\,\left[\mathcal{R}_\text{b}+\frac{12}{R^2}\right],
\end{equation}

\begin{figure}
  \includegraphics[width=3.5 in]{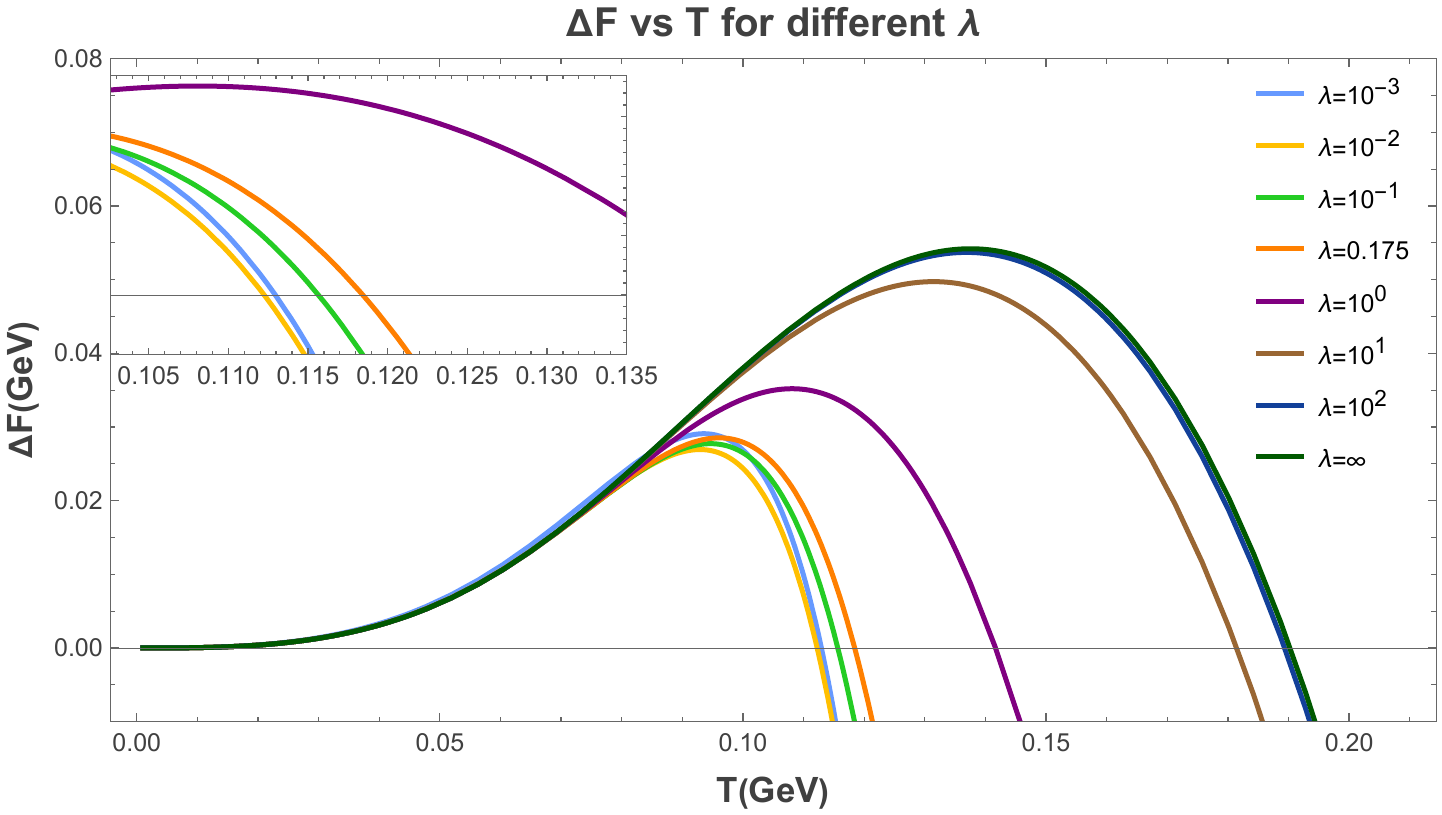}
\caption{Hawking-Page transition for different isospectral probe dilaton fields $\Phi_\lambda(z)$.}
\label{fig:HP}
\end{figure}

\noindent where the subscript $b$ labels the background metric (either thermal AdS or AdS–Schwarzschild), $\mathcal{R}_\text{b}$ is the corresponding Ricci scalar, and $G_5$ is the five‑dimensional Newton constant. The free energy difference $\Delta F = I_{\text{Th}}/T - I_{\text{AdS-Schw}}/T$ determines the preferred phase. The critical deconfinement temperature $T_c^{\text{(HP)}}$ is the temperature at which $\Delta F$ changes sign. As shown in Fig.~\ref{fig:HP}, $T_c^{\text{(HP)}}$ increases monotonically with the isospectral parameter $\lambda$, in agreement with the observations for scalar glueballs in Ref.~\cite{Afonin:2018era}.

This behavior is opposite to that of the $\rho$ meson melting temperature $T_m$ extracted from spectral functions, which decreases with $\lambda$ (see Fig. \ref{fig:c}). The opposite trends are not contradictory: the two critical temperatures probe different sectors of the theory. The Hawking–Page transition depends on the dilaton through the overall factor $e^{-\Phi_\lambda(z)}$ in the gravitational action, which modifies the free energy difference. In contrast, the melting temperature is obtained from the spectral function of a specific hadronic mode, whose equation of motion involves the dilaton in a distinct way (via the function $B_\lambda(z)$ in Eq.'s  \eqref{bulk-to-boundary} and \eqref{Green-Th}). The isospectral transformation, therefore, affects the two critical temperatures in opposite directions.

For the vector meson case studied here, increasing $\lambda$ (which lowers the ground‑state decay constant $f_1$) raises the deconfinement temperature from the Hawking–Page analysis while lowering the melting temperature from the spectral function. This underscores the importance of specifying which observable is used to define the \emph{critical temperature} in holographic models.\par

\subsection{The special point $\lambda=0.175$ in the isospectral family}
The AdS/QCD softwall model \cite{Karch:2006pv} has been shown to predict small melting (dissociation) temperatures \cite{Vega:2017dbt}. As we mentioned earlier, this fact is a consequence of the ill-defined EM decay constants $f_n$, which, for the vector softwall model, are degenerate in the zero temperature case \cite{Braga:2015jca}.

The isospectral transformation \emph{fixes} the ground state decay constant when isospectrality comes to play in the scene. It was found that choosing $\lambda =0.175$ defines a probe dilaton that reproduces $f_{\rho(770)}=226$ MeV in good agreement with experimental data \cite{MartinContreras:2023eft}. Here we prove the consistency of such a dilaton by looking at the behavior of the $\rho$ meson spectral peak at finite temperature. For simplicity, we fix $\mu=0$ for the present discussion.

\begin{figure*}
 
  \includegraphics[width=3.4 in]{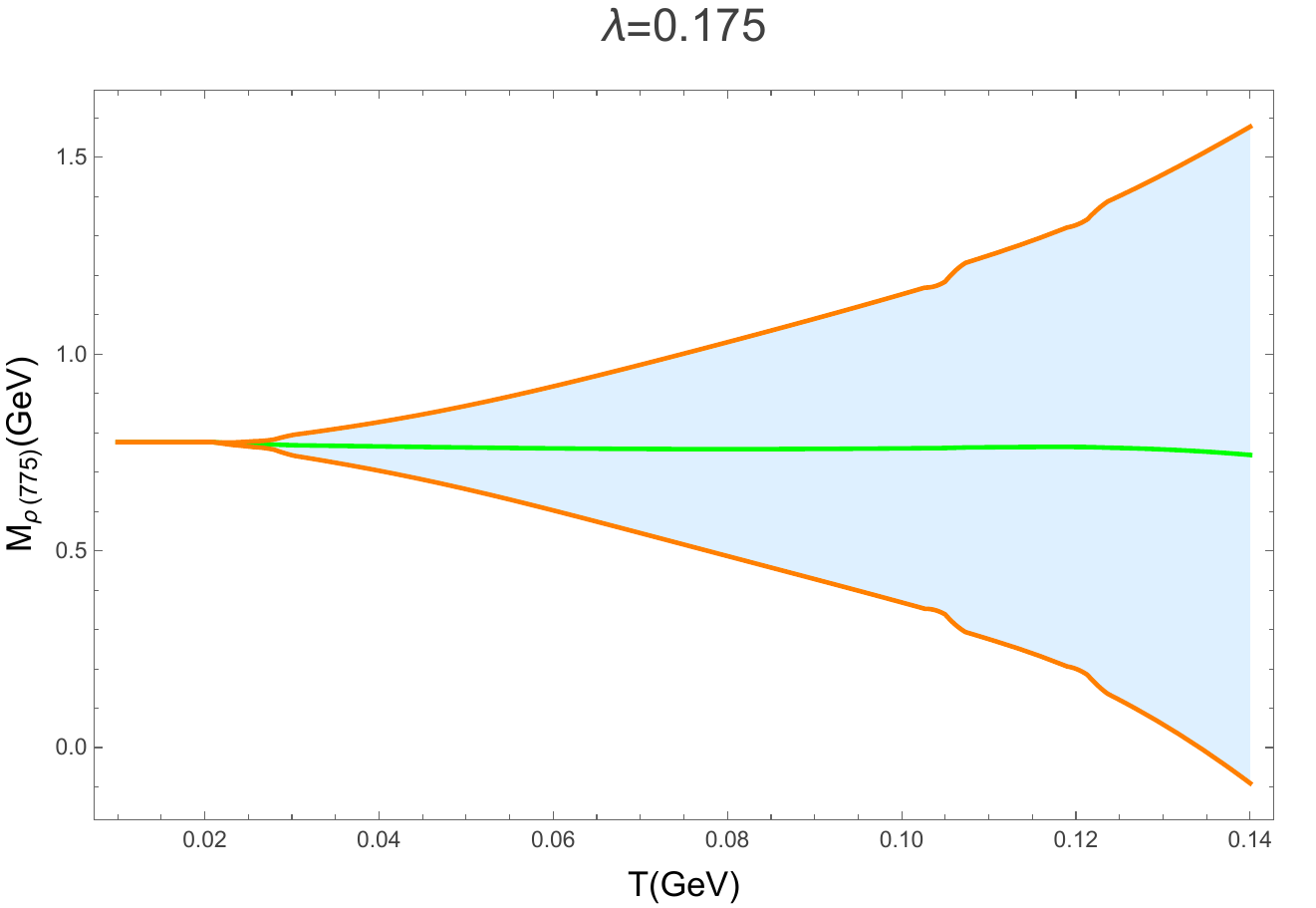}
   \includegraphics[width=3.4 in]{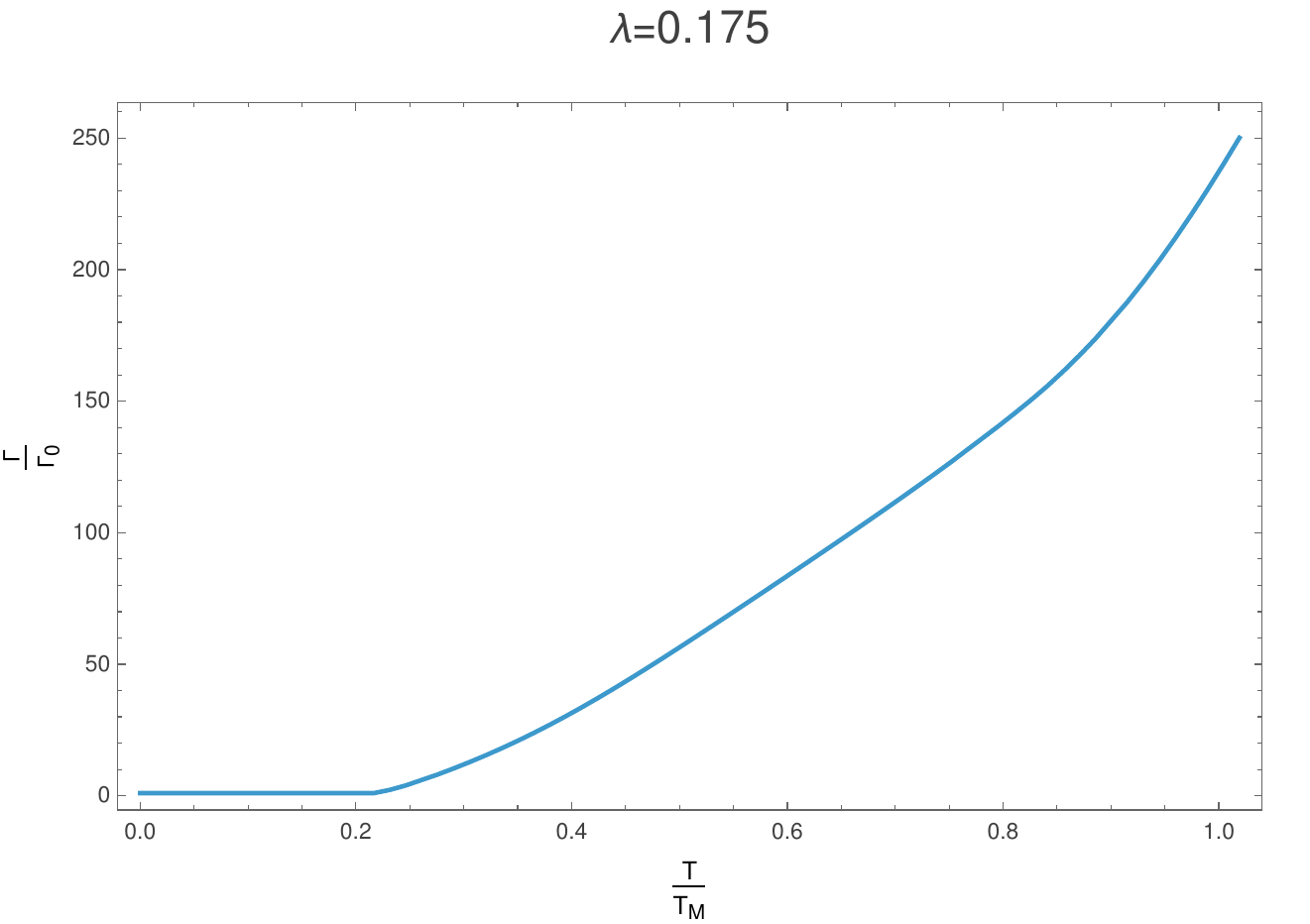}

\caption{Left panel shows thermal mass of the $\rho$ meson ground state and the corresponding thermal width for $\lambda=0.175$. For this choice of $\lambda$, the critical temperature for the meson dissociation is $157$ MeV. In the right panel, we depict the $\rho$ meson branching ratio as a function of temperature for  $\lambda=0.175$. The smooth growth of the branching ratio near the critical temperature $T_m$ is compatible with a crossover transition between confined and deconfined matter.}
\label{fig:0175}
\end{figure*}

In Figure \ref{fig:0175} we summarize the evolution of the $\rho$ meson spectral peak, including its thermal mass (left panel) and thermal width (right panel), from low temperatures up to the melting temperature. We found that the corresponding melting temperature is $T_m = 157$ MeV. We observe a small decrease in the $\rho$ meson thermal mass near the critical temperature. The critical temperature and the behavior of the thermal mass agree with the results obtained using thermal sum rules presented in \cite{Ayala:2012ch}. These differ from the results obtained using chiral perturbation theory \cite{Dominguez:1992dw}, which predicts a monotonic increase of the $\rho$ meson mass with temperature.\par

In the right panel of Fig. \ref{fig:0175}, we plot the thermal width for the $\rho(770)$ meson peak as a function of the temperature. In this case, we found qualitative agreement with the results of thermal sum rules: the branch ratio increases monotonically with temperature. However, there is a crucial difference. In \cite{Ayala:2012ch}, the dependence of $\Gamma(T)$ is implemented by means of an \emph{a priori} phenomenological fit, motivated by Finite Energy QCD Sum Rules (FESR). When approaching $T_m$, the FESR ceases to have solutions with the assumed Breit-Wigner-like spectral function. The perturbative threshold $s_0(T)$ vanishes at $T_m$, indicating that \emph{there are no hadronic degrees of freedom to describe the system in the thermal medium}. Therefore, it implies that above $s_0(T)$, the quarks and gluons described by perturbative QCD are expected to represent the system dynamics accurately. In plain words, near $T_m$, there is no longer a quasiparticle spectral peak.\par
 
This FESR scenario represents a first-order phase transition for quark deconfinement. On the other hand, the holographic results we obtain using the isospectral transformation do not assume any particular scenario for the phase transition \emph{a priori}. As a result, we find that the holographic approach is compatible with a crossover transition from confinement to deconfinement.

\section{Excited states}
\label{sec:6}
Concerning the spectral properties of the $\rho$ meson trajectory, the isospectral transformation adjusts only the ground-state decay constant, keeping all other quantities fixed. As we discussed in Section \ref{sec:5}, the changes in the ground state decay constant reflect in different critical temperatures for the dissociation of the $\rho$ meson in the thermal medium. From the current-current correlator, we expect that the spectral properties of the excited states, defined by the poles and residues, are completely invariant under the isospectral transformation. There should be no changes in the quasiparticle peaks of the spectral functions at finite temperature or chemical potential when obtained using different probe dilatons from the same isospectral family. However, this is not the case. 

\begin{figure*}
 
  \includegraphics[width=4.0 in]{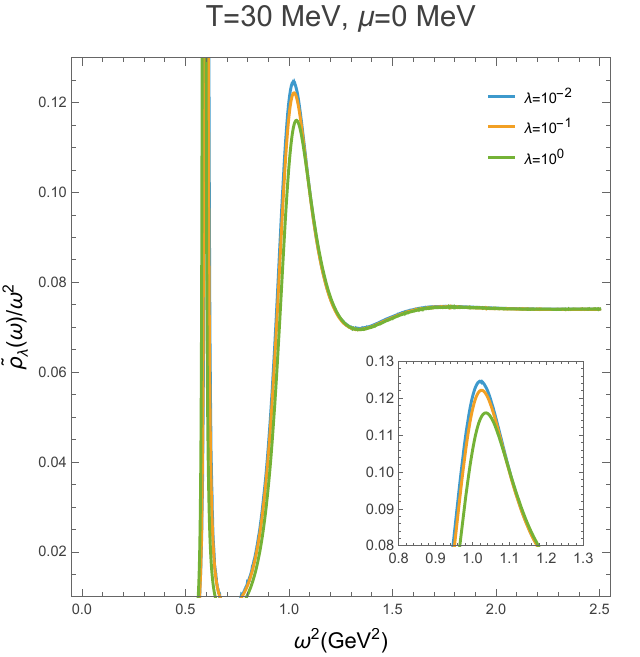}

\caption{Spectral peak of the first excited state of $\rho$ meson is sensitive on the isospectral parameter.}
\label{fig:excited}
\end{figure*}

In Figure \ref{fig:excited}, we see the first excited state spectral peaks of the $\rho$ meson for different dilatons in the same isospectral family at $30$ MeV and zero chemical potential. We plot the spectral functions for three different values of the isospectral parameter $\lambda$. The choice of $T=30$ MeV is such that the peak of the first-excited state is still pronounced so that we can analyze it.

We observed that, for smaller values of $\lambda$ associated with larger decay constants for the ground state, the quasiparticle peaks for the excited states are slightly more pronounced. This effect is more evident in the first excited state. For higher excitations, it is still noticeable that the peaks are slightly more pronounced for smaller values of $\lambda$. However, for higher radial excitations, the effect is suppressed by the thermal instability of such states.\par

The behavior of the peaks corresponding to the excited states is unexpected from the perspective of zero-temperature isospectrality. However, this outcome can be understood from the effect of the probe dilaton $\Phi_\lambda(z)$ over the thermal Green's function \eqref{Green-Th}, that scales the spectral function profile with $\lambda$.\par

\section{Conclusions}
\label{sec:7}

We have studied the thermal and dense behavior of $\rho$ mesons within the isospectral family of the softwall AdS/QCD model. By computing the spectral function $\rho_\lambda(\omega)$ at finite temperature and chemical potential for different values of the isospectral parameter $\lambda$, we were able to vary the ground-state electromagnetic decay constant $f_1$ while keeping the entire mass spectrum unchanged. This unique property allowed us to isolate the role of $f_1$ in the melting process. Our main finding is a monotonic increase of the $\rho(770)$ melting temperature $T_m$ with $f_1$, indicating that a more compact $q\bar{q}$ configuration (larger $f_1$) is more resistant to dissociation in a thermal medium. The relationship is nonlinear but clearly positive, and it persists, albeit with suppressed amplitude, for the first radial excitation. Higher excitations become increasingly insensitive to the value of $f_1$, a pattern that reflects the growing dominance of the thermal continuum over the discrete bound-state structure.

When the isospectral parameter is set to $\lambda = 0.175$, the model reproduces the experimental decay constant of the $\rho(770)$ meson ($f_1 \approx 226$ MeV). The corresponding spectral function yields a melting temperature $T_m = 157$ MeV, in good agreement with lattice QCD estimates of the deconfinement crossover. The thermal mass $M_\rho(T)$ remains nearly constant up to $T \sim 150$~MeV and then decreases slightly, while the thermal width grows monotonically. The smooth disappearance of the quasiparticle peak near $T_m$ is consistent with a crossover transition, in contrast to a first-order melting scenario. The same holographic setup, when analyzed via the Hawking–Page transition, gives a deconfinement temperature of about $118$ MeV (depending on $\lambda$), which is lower than the melting temperature. This discrepancy is not a contradiction: the two observables probe different aspects of the theory: the free energy of the bulk background versus the response of a specific hadronic excitation. Therefore, their $\lambda$ dependences are opposite, reflecting the distinct forms in which the isospectral probe dilaton enters the gravitational action versus the matter field action.\par

From a broader perspective, our work demonstrates that the isospectral transformation is a powerful tool for bottom-up holographic QCD. It enables a systematic separation of mass parameters from decay constants, allowing one to fine‑tune the latter without altering the Regge trajectory. The observed small deviations from strict isospectrality in thermal masses (expected because the transformation is defined for the vacuum Schrödinger equation, not for the thermal wave equation) are quantitatively under control and do not undermine the utility of the construction.

The isospectral paradigm thus opens a controlled avenue for improving holographic predictions of in‑medium hadron properties while preserving the zero‑temperature spectroscopy.

\begin{acknowledgments}
M. A. Martin Contreras would like to acknowledge the financial support provided by the National Natural Science Foundation of China (NSFC) under grant No. 12350410371. A. Vega is partially supported by the Centro de Física Teórica de Valparaíso (CeFiTeV) and by FONDECYT (Chile) under grant No. 1251106. Saulo Diles thanks the Conselho Nacional de Desenvolvimento Científico e Tecnológico (CNPq), Brazil, for Grant No. 406875/2023-5.
\end{acknowledgments}

\bibliography{apssamp}

\end{document}